\documentclass[prb,showpacs,twocolumn]{revtex4}
\usepackage{setspace}
\usepackage{graphicx}

\begin{document}

\title{Non-ohmic spin transport in n-type doped silicon}
\author{Hyuk-Jae Jang}
\author{Jing Xu}
\author{Jing Li}
\author{Biqin Huang}
\author{Ian Appelbaum}
\altaffiliation{Present address: Physics Department, University of Maryland (appeli@umd.edu)}
\affiliation{Department of Electrical and Computer Engineering,
University of Delaware, Newark, Delaware, 19716}
\begin{abstract}
We demonstrate the injection and transport of spin-polarized electrons through n-type doped silicon with in-plane spin-valve and perpendicular magnetic field spin precession and dephasing (``Hanle effect'') measurements. A voltage applied across the transport layer is used to vary the confinement potential caused by conduction band-bending and control the dominant transport mechanism between drift and diffusion. By modeling transport in this device with a Monte-Carlo scheme, we simulate the observed spin polarization and Hanle features, showing that the average transit time across the short Si transport layer can be controlled over 4 orders of magnitude with applied voltage. As a result, this modeling allows inference of a long electron spin lifetime, despite the short transit length.
\end{abstract}


\maketitle
\newpage

\section{INTRODUCTION}

It has been a longstanding goal in semiconductor spintronics to inject, transport, manipulate, and detect spin-polarized carriers in silicon-based devices.\cite{zutic,fabian,lyon} Despite great success in the field over the past ten years using other semiconductors such as GaAs,\cite{awschalom,jiang,crooker,lou,dery,holub} the goal of achieving the same with Si has been reached only recently, using all-electrical hot-electron methods with undoped single-crystal silicon transport layers.\cite{appelbaum} Later, spin injection into silicon was realized as well in an epitaxially-grown silicon n-i-p diode structure using circular polarization analysis of weak electroluminescence spectra for spin detection across a transport layer of 80-140nm.\cite{jonker} Although our previous studies demonstrate electron spin manipulation in undoped silicon - even over a very long distance (350 microns)\cite{huang, OBLIQUE, DEPHASING} - it is necessary to investigate magnetic- and electric-field control of electron spin in {\it doped} silicon for integration of spintronics into present-day silicon-based microelectronic technology, where impurity doping plays a critical role.

In this report, we present spin injection, transport and detection in an n-type doped silicon device using our all-electrical methods. Unlike previous studies with undoped Si, the presence of ionized impurities in the depletion regions of these doped transport layers gives rise to conduction band bending that for sufficient biasing conditions confines injected electrons for long dwell times. By modeling transport with drift and diffusion in the inhomogeneous electric fields provided by the band bending with a Monte-Carlo method, we simulate both spin precession and spin decay, showing that the transit time distribution of spin-polarized electrons can be controlled over a very wide range with an applied voltage, and can yield a measurement of spin lifetime.

\section{EXPERIMENTAL METHODS}

\begin{figure}
\centering
\includegraphics[width=8.5cm,height=3.5cm]{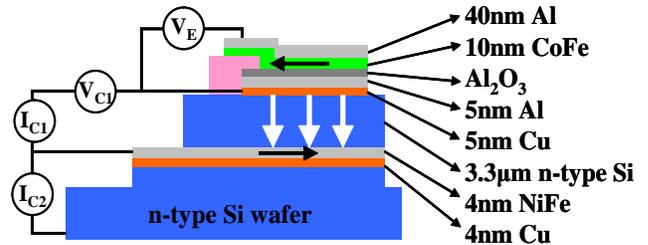}
\caption{Schematic side view of n-type doped Si spin transport device and illustration of the components and contacts for electrical injection and detection.}
\label{FIG1}
\end{figure}

 Fig. \ref{FIG1} illustrates the structure of our device. Fabrication consists of ultra-high vacuum metal film wafer bonding to assemble a semiconductor-metal-semiconductor hot-electron spin detector; a silicon-on insulator (SOI) wafer including a 3.3$\mu$m single-crystal (100) nominally 1-20 $\Omega\cdot$cm phosphorus-doped n-type silicon spin transport layer is bonded to an n-type bulk silicon collector wafer with a Ni$_{80}$Fe$_{20}$ (4nm)/ Cu (4nm) bilayer. Conventional wet-etching techniques expose the SOI device layer, onto which a ferromagnetic-emitter tunnel junction hot-electron spin injector is built. The final device structure is Al (40nm)/Co$_{84}$Fe$_{16}$ (10nm)/Al$_{2}$O$_{3}$/Al (5nm)/Cu (5nm)/n-Si (3.3 $\mu$m)/Ni$_{80}$Fe$_{20}$ (4nm)/Cu (4nm)/n-Si substrate, as displayed in Fig. 1. Further details on fabrication of similar devices can be found in previous reports\cite{appelbaum,huang,zhao}.

An applied emitter voltage $V_{E}$ on the tunnel junction (larger than the Cu/n-Si injector Schottky barrier) injects hot electrons tunneling from the ferromagnetic Co$_{84}$Fe$_{16}$ cathode through the thin-film Al/Cu anode base and into the doped silicon transport layer conduction band. The first collector voltage ($V_{C1}$) controls the voltage drop across the transport layer and modifies the spatially nonlinear conduction band potential energy. Electrons escaping the transport layer are ejected over a Schottky barrier at the detector side into hot-electron states in a buried Ni$_{80}$Fe$_{20}$ thin film. The final spin polarization is detected by measuring the ballistic component of this hot electron current (second collector current, $I_{C2}$) in the n-type Si wafer below; spin-dependent scattering in the ferromagnetic Ni$_{80}$Fe$_{20}$ makes this current dependent on the projection of final spin angle on the Ni$_{80}$Fe$_{20}$ detector magnetization.

\section{EXPERIMENTAL RESULTS}

\begin{figure}
\centering
\includegraphics[width=8cm,height=7cm]{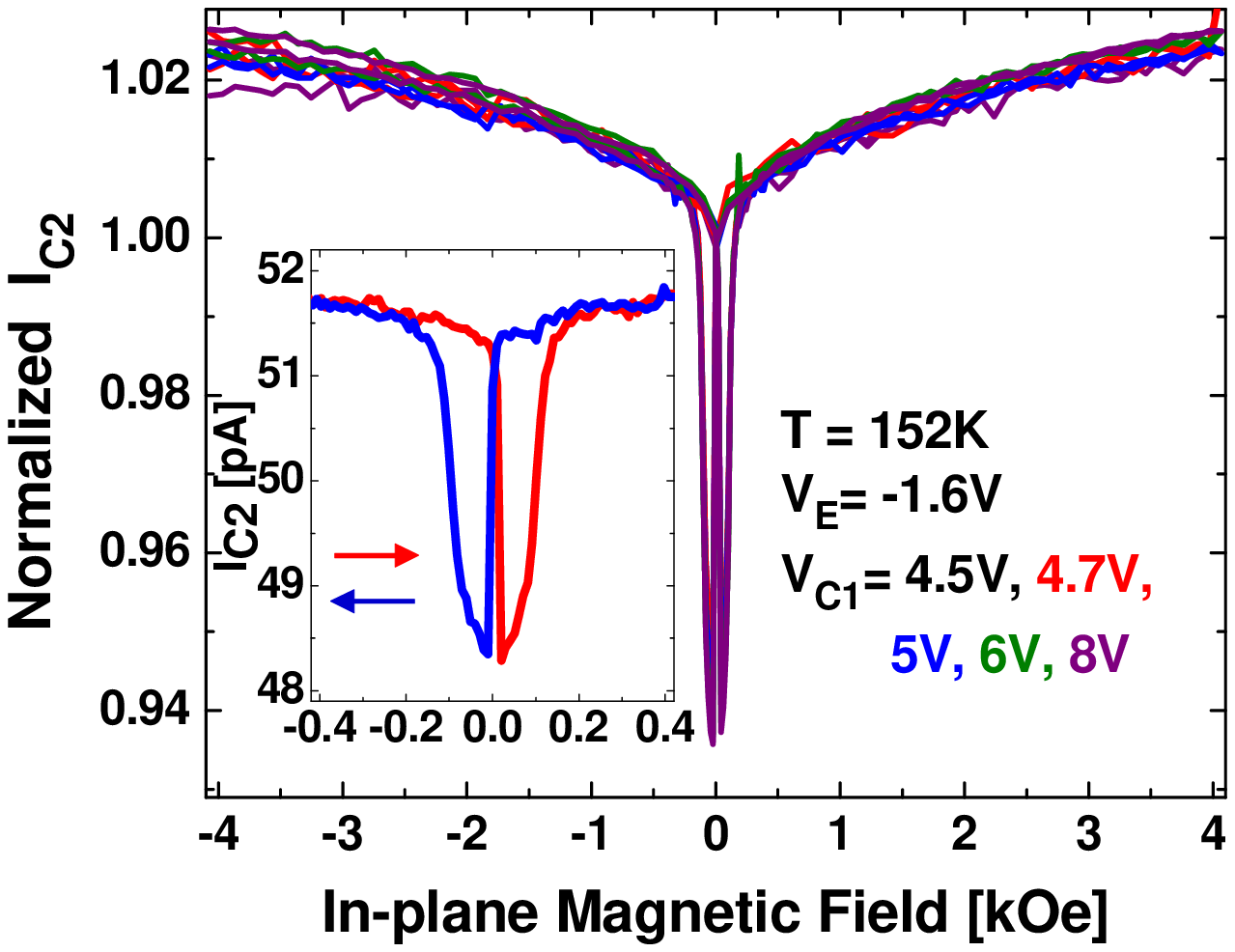}
\caption{Spin valve effect in doped Si spin-transport devices. Measurement was done at T = 152K and V$_{E}$ = -1.6V, with different $V_{C1}$ values applied as indicated in the plot. Inset: Data measured with $V_{C1}$ = 5V plotted over a smaller field range. Field sweep directions are indicated by red (increasing) and blue (decreasing) arrows.}
\label{FIG2}
\end{figure}

The spin-detection current $I_{C2}$ was first measured with an external magnetic field parallel to the device plane. A spin-valve effect, resulting from the different in-plane coercive fields of injector and detector ferromagnetic layers, is displayed in Fig. \ref{FIG2}. The measurements were done with $V_{E}$ = -1.6V applied, using different values of $V_{C1}$ between 4.5V and 8V at temperature $T$ = 152K. Because of the $I_{C2}$-$V_{C1}$ dependence, we normalize the data for comparison between different $V_{C1}$ values. After this normalization, it can be seen that the measurement is only weakly dependent on accelerating voltage $V_{C1}$ over this range.

The in-plane magnetic field was swept between -4 kOe to +4 kOe for this measurement. Since the coercive fields of both ferromagnetic (FM) layers are smaller than 200 Oe, the data obtained from the $V_{C1}$ = 5V measurement is magnified in the inset of Fig. 2 and the field sweep direction is specified by correspondingly colored arrows. When the in-plane magnetic field reaches approximately +20 Oe from the negative saturation field (below -300 Oe), the Ni$_{80}$Fe$_{20}$ layer switches its magnetization, causing an anti-parallel (AP) configuration in the two FM layers, which lowers the $I_{C2}$ current relative to a parallel (P) configuration, because in this case spin ``up'' is injected, but spin ``down'' is detected. If the magnetic field increases further, the Co$_{84}$Fe$_{16}$ layer reverses magnetization, resulting in a P configuration and restoration of the higher $I_{C2}$. This happens as well in the opposite sweeping field direction due to the symmetric but hysteretic coercive fields of each FM layer. The magnetocurrent (MC) ratio ($I_{C2}$$^{P}$\textendash \space $I_{C2}$$^{AP}$)/$I_{C2}$$^{AP}$ calculated from the spin-valve plot, where the superscripts refer to P and AP magnetization configurations in the two FM layers, is approximately 6\%. As the magnetic field reaches up to $\pm$4 kOe after the magnetization reversal of both FM layers, $I_{C2}$ monotonically rises because of domain magnetization saturation in the direction of the external field.

To unambiguously confirm spin transport through the doped silicon layer, we have performed measurements of $I_{C2}$ in an external magnetic field perpendicular to the device plane, which allows us to examine spin precession and dephasing (Hanle effect) during transport.\cite{appelbaum,huang,johnson,jedema,crowell,huang2, zhao, DEPHASING, monsma, OBLIQUE} Depending on the magnitude of the applied magnetic field and the transit time (subject to drift and diffusion through the conduction band from injector to detector), the polarized electron spin (initially parallel to the injector FM layer magnetization) can arrive at the detector having rotated through precession angle $\theta=\tau g \mu_B B / \hbar$, where  $\tau$ is the transit time, $B$ is the magnetic field, $g$ is the electron spin g-factor, $\mu$$_{B}$ is the Bohr magneton, and $\hbar$ is the reduced Planck constant.

\begin{figure}
\centering
\includegraphics[width=8.5cm,height=7.5cm]{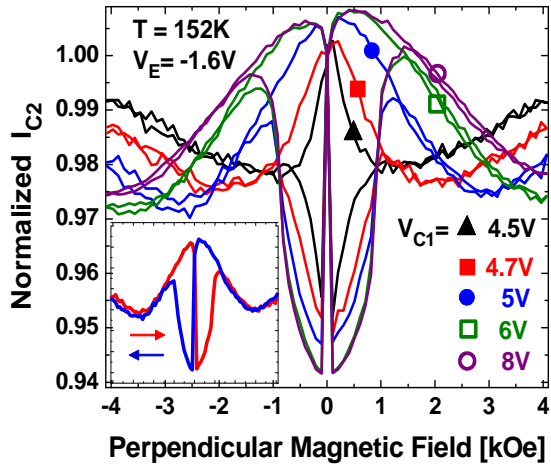}
\caption{\label{FIG3} Electron spin precession in doped Si spin-transport devices. Emitter voltage and temperature are same as in Fig. \ref{FIG2}. Minima, corresponding to $\pi$ rad precession angle appear at higher magnetic field as $V_{C1}$ increases and transit time decreases. Inset: data measured with $V_{C1}$ = 5V. Field sweep directions are indicated by red (increasing) and blue (decreasing).}
\end{figure}

Our measurements in a perpendicular magnetic field, using the same experimental conditions as were applied in the spin-valve effect measurement (V$_{E}$ = -1.6V and T = 152K), are shown in Fig. \ref{FIG3} for the same varied values of $V_{C1}$ as in Fig. \ref{FIG2}. The measured $I_{C2}$ was normalized for data comparison at different accelerating voltages $V_{C1}$, as in the spin-valve effect experiment. Again, the inset of Fig. \ref{FIG3} shows the data for $V_{C1}$ = 5V with magnetic field sweep directions indicated by correspondingly colored arrows. When a perpendicular magnetic field sweeps from -4 kOe (or from +4 kOe), $I_{C2}$ exhibits a minimum before the field reaches 0 Oe and then it suddenly drops and slowly moves up between 0 and +1kOe. The former minima is induced by a full spin flip due to spin precession (average $\pi$ rad rotation) during transport through the doped silicon layer, and the latter is induced by the in-plane magnetization switching of the two FM layers by a residual in-plane component of the largely perpendicular magnetic field, causing an antiparallel injector/detector magnetization configuration and reduction in signal as seen in previously-discussed in-plane spin-valve measurements. This argument is further upheld by changing $V_{C1}$; minima attributed to precession appear at higher magnitude of applied perpendicular magnetic field as $V_{C1}$ increases due to the shorter transit time, while the FM switching fields clearly do not change.\cite{monsma}

\section{EXPERIMENTAL ANALYSIS}

Average spin transit times $\tau$ on the order of 45 - 180ps can be determined from the magnetic field values at $\pi$ rad precession minima $B$$_{\pi}$ in Fig. 3 ($\sim$ 1 kOe \textendash \space 4 kOe) using\cite{monsma}  $\tau=h/2g\mu_B B_{\pi}$, where $h$ is the Planck constant. Correlating spin polarization from spin-valve measurement to these transit times can, in principle, be used to determine spin lifetime. However, these transit times are very short, so direct correlation as in Ref. \cite{huang} is unable to independently determine the (long) spin lifetime of conduction electrons in doped Si. We have previously measured spin lifetime of 73ns at similar temperature using a 350 micron-thick undoped Si transport layer device; this lifetime increases to over 500ns at 60K.\cite{huang}

\begin{figure}
\centering
\includegraphics[width=8cm,height=7.5cm]{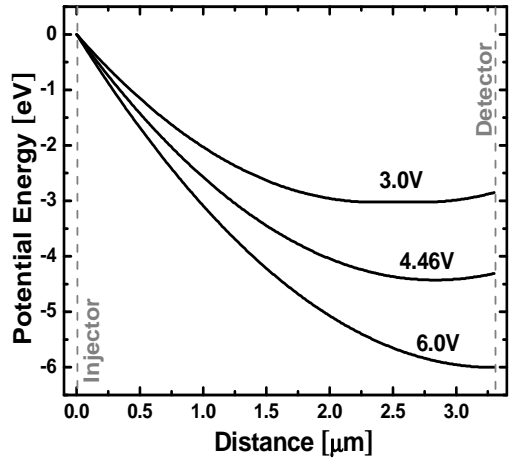}
\caption{Depletion-approximation conduction-band diagrams of the doped Si spin transport layer with injector-detector voltage drop of (a) 3.0V (where the transport is dominated by diffusion against the electric field at the detector side); (b) 4.46V (where the bias is enough to eliminate the neutral region and fully deplete the transport layer); and (b) 6.0V (where the potential well has been eliminated and transport is dominated by drift in the unipolar electric field).}
\label{BANDDIAGRAM}
\end{figure}

In the undoped silicon transport layers used in previous works,\cite{appelbaum, huang, monsma, DEPHASING, OBLIQUE} the Schottky depletion region was much larger than the layer thickness. Therefore, the conduction band was linear, resulting in a spatially constant induced electric field, and relatively ``ohmic'' spin transport where the spin transit time was inversely proportional to the injector-detector voltage drop. In this work, however, carrier depletion of the doped silicon due to Schottky contacts and the resulting space-charge from ionized impurities causes a nonlinear conduction band that can have a potential energy minimum between depletion regions unless the voltage drop is very large. Since injected electrons may sit in this potential well for a long time before escaping over the detector barrier, their spins will depolarize and the observed MC ratio will be suppressed.

To significantly reduce this dwell time, an accelerating voltage (induced by applied voltage $V_{C1}$, which adds to approximately 0.3V of the applied emitter voltage due to resistive tunnel junction electrodes\cite{monsma}) can be used to alter the confining potential energy. In particular, for sufficient voltage the confining potential can be eliminated. It is therefore expected that the spin signal is strongly sensitive to applied voltage and ``non-ohmic'' spin transport results.

\section{MODEL}

Modeling this non-ohmic behavior is necessary. In previous works using undoped Si transport layers where the electric field is constant from injector to detector, a modeling technique using the arrival-time distribution given by the Green's function solution to the drift-diffusion equation can be easily implemented.\cite{crowell,huang, DEPHASING, OBLIQUE} However, the electric field in these doped Si devices is highly inhomogeneous, making it difficult to implement the standard method here because the drift velocity is spatially dependent, requiring Green's function solution of a nonlinear partial differential equation. In general, this procedure is non-trivial.

\begin{figure}
\centering
\includegraphics[width=7.5cm,height=15cm]{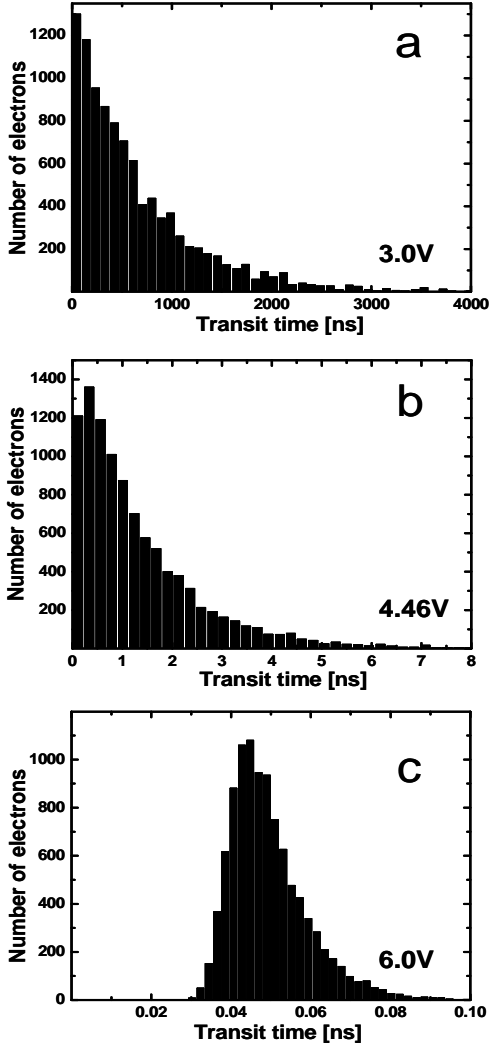}
\caption{\label{DISTRIBUTIONS} Monte-Carlo simulated transit-time distributions for injector-detector voltage drops of (a) 3.0V; (b) 4.46V; and (c) 6.0V. Note timescale changes over 4 orders of magnitude from (a) to (c). }
\end{figure}

\begin{figure}
\centering
\includegraphics[width=7.5cm,height=7.5cm]{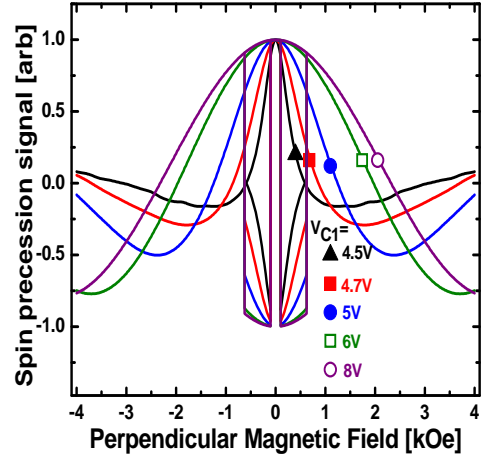}
\caption{\label{HANLE} Results of Hanle effect simulation, Eq. (\ref{MCHANLE}). Similar to the experimental results in Fig. \ref{FIG3}, minima corresponding to $\pi$ rad precession angle appear at higher magnetic field as $V_{C1}$ increases and transit time decreases. Magnetization switching is simulated by signal sign reversal between the coercive field values of the injector and detector ferromagnets.}
\end{figure}

\begin{figure}
\centering
\includegraphics[width=7.5cm,height=12cm]{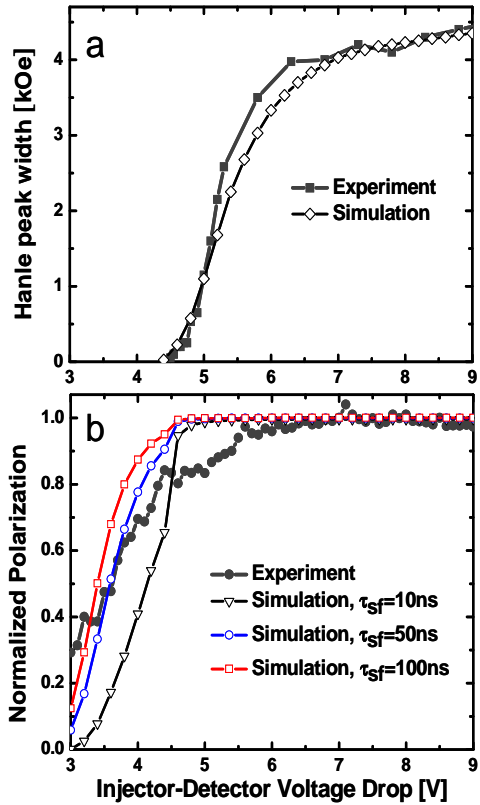}
\caption{\label{HANLEWIDTHPOL} Comparison of experimental and Monte-Carlo simulated voltage dependence of (a) Hanle peak full width at half maximum; and (b) spin polarization $\mathcal{P}=(I_{C2}^P-I_{C2}^{AP})/(I_{C2}^P+I_{C2}^{AP})$ at 152K. The simulation shown in (a) is insensitive to a choice of $\tau_{sf}$ over the range 10-100ns. }
\end{figure}

To overcome this problem and simulate spin transport behavior in these doped devices, we use a Monte-Carlo technique which translates electrons a distance $v(x) \Delta t$ (due to drift), and $\pm \sqrt{2D\Delta t}$ (due to diffusion) in a timestep $\Delta t$, where $v(x)$ is the drift velocity at the position $x$ and $D$ is the diffusion constant. (The sign on the latter expression is randomly chosen to simulate the stochastic nature of 1-dimensional diffusion.) 

The spatially-dependent electric field is calculated within the depletion approximation. Using a doping density of $7.2\times 10^{14}$ cm$^{-3}$, injector Schottky barrier height of 0.6eV (for Cu/Si) and detector Schottky barrier height of 0.75eV (for NiFe/Si) results in a band diagram whose dependence on injector-detector voltage drop is shown in Fig. \ref{BANDDIAGRAM}. This figure illustrates that the voltage drop across the Si transport layer can be used to alter the dominant transport mode: at low bias a wide neutral region exists between depletion regions and electrons must diffuse {\it against an electric field} to escape to the detector, whereas for biases greater than 6V, the potential minimum is annihilated by the boundary so the internal electric field carries electrons toward the detector everywhere and drift is expected to dominate. 

A realistic empirical mobility model using Eq. 10 from Ref. \cite{MOBILITY} is used to evaluate $v(E)$. The diffusion coefficient at each point in space is then calculated from the Einstein relation $D(x)=\mu(x) k_B T / q$, where mobility $\mu(x)=v(x)/E(x)$ and $E(x)$ is electric field. We simulate transport for $10^4$ electrons at each value of injector-detector voltage drop and the arrival time at the detector for each is recorded. The distribution of arrival times $f(t, V)$ is constructed from a histogram of this data and are used to calculate the expected output due to spin precession in a perpendicular magnetic field (Hanle effect):

\begin{equation}\label{MCHANLE}
I_{C2}(V)=\int_0^\infty f(t,V)(\cos^{2}{\theta}\cos{\omega t}+\sin^{2}{\theta})e^{-t/\tau_{sf}}dt.
\end{equation}

\noindent where $\theta$ is the angle between the injector/detector magnetization and the device plane, $\tau_{sf}$ is effective spin lifetime and the spin precession angular frequency $\omega=g\mu_B B/\hbar$. The tilting angle $\theta$ is caused by the external magnetic field partially overcoming the finite geometric anisotropy of the magnetic thin films\cite{jedema} and is necessary to correctly model the experimental results. As in Ref. \cite{valenzuela}, we use $\theta=\sin^{-1}(B/15.5kOe)$. 

In addition, the final spin polarization after transport can be calculated from

\begin{equation}\label{MCPOL}
\mathcal{P}(V)=\int_0^\infty f(t,V)e^{-t/\tau_{sf}}dt.
\end{equation}

\section{MODEL RESULTS}

As can be seen in Fig. \ref{BANDDIAGRAM}, an electric field opposing transport to the detector is present at low voltage. Electrons must therefore diffuse against this electric field to escape the confining potential in the bulk of the Si transport layer. Under these conditions of diffusion-dominated transport, the arrival-time distribution has a very wide exponential shape with average transit time of approximately 500ns, as shown in Fig. \ref{DISTRIBUTIONS} (a). Although the width of the distribution can be reduced significantly by increasing the voltage drop to the point where the Si transport layer is fully depleted as shown in Fig. \ref{DISTRIBUTIONS} (b), the confining electric field remains and the exponential shape is maintained. This indicates that diffusion is still strong.

For sufficiently high voltage drops, the potential energy minimum is annihilated by the detector boundary as indicated in Fig. \ref{BANDDIAGRAM} and drift-dominated transport occurs. This is reflected in the gaussian-like shape of the distribution in Fig. \ref{DISTRIBUTIONS} (c) for a voltage drop of 6V. At this voltage, the average transit time is only approximately 50ps, consistent with the analysis of experimental Hanle effect measurements. Therefore, as a result of our Monte-Carlo modeling we see that the average electron transit time in our doped Si spin transport devices can be controlled over approximately 4 orders of magnitude by changing the injector-detector voltage drop by only several volts (from 3V to 6V).

Using Eq. (\ref{MCHANLE}), we simulate the Hanle effect in our devices using $\tau_{sf}=100$ns (choice of this value will be discussed later). Fig. \ref{HANLE} shows Hanle effect simulations for voltages corresponding to the same $V_{C1}$ values as in Fig. \ref{FIG3} (again, a shift of 0.3V due to a portion of the emitter bias dropping across the resistive tunnel junction base\cite{monsma} is accounted for to make a direct comparison) in wide agreement to those experiments. In particular, the qualitative shape and precession minima positions are well modeled.

The most salient feature of the Hanle effect simulation is the magnetic-field width of the central (zero precession-angle) peak, plotted as a function of injector-detector voltage drop in Fig. \ref{HANLEWIDTHPOL}(a) and compared to the experimental values. Note that the width is constant for voltages greater than 6V (due to drift velocity saturation at high electric field in Si), and the presence of a threshold near 5V (due to appreciable lowering of the confining potential barrier at the detector side of the transit layer once full-depletion occurs at approximately that voltage). This sudden collapse of the Hanle peak width is not seen in the voltage dependence of spin precession measurements using undoped drift-dominated spin transport devices.

In Fig. \ref{HANLEWIDTHPOL} (b), we show the voltage dependence of the measured spin polarization $\mathcal{P}=(I_{C2}^P-I_{C2}^{AP})/(I_{C2}^P+I_{C2}^{AP})$ from experimental data using in-plane magnetic field spectroscopy as described in Fig. \ref{FIG2}. Again, a threshold is seen in the experimental data. However, the position of the spin polarization threshold in Fig. \ref{HANLEWIDTHPOL} (b) near 3.5V is at much smaller bias voltage as compared to the Hanle width collapse threshold near 5V shown in Fig \ref{HANLEWIDTHPOL}(a). This indicates that the electrons maintain their spin despite a long dwell time which causes strong spin dephasing in the confining conduction band potential minimum at low voltages. Comparing this behavior to the model results from Eq. (\ref{MCPOL}) with different values of spin lifetime $\tau_{sf}$ shows that this discrepancy in threshold position is consistent with a long spin lifetime of 10-100ns. This can be compared to a spin lifetime of approximately 73ns measured in undoped Si at the same temperature using a different technique.\cite{huang}

\section{CONCLUSION}

In summary, we have demonstrated spin transport through n-type doped Si. Using a Monte-Carlo algorithm to model drift and diffusion, we simulated electron transport through the inhomogeneous internal electric field and make quantitative comparisons to experimental values of spin polarization and Hanle peak width without any free fitting parameters. Analysis of the arrival-time distribution indicates that in doped transport layers, the spin-polarized electron transit time can be controlled over several orders of magnitude with applied voltage. The resulting non-ohmic behavior seen here is in contrast to spin transport measurements using undoped silicon transport layers, and is expected to influence future semiconductor spintronic device designs utilizing current-sensing spin detection methods in n-type doped semiconductors.

\vspace{1 pc}

The authors gratefully acknowledge helpful discussions with D. Weile and support from DARPA/MTO, the Office of Naval Research, and NSF.

\end{document}